\documentclass[a4paper,12pt]{article}
\usepackage{amsfonts}
\usepackage{latexsym}
\usepackage{amsmath}
\usepackage{amssymb}
\usepackage{amssymb}
\usepackage{slashed}
\usepackage{upgreek}
\usepackage{mathrsfs}
\usepackage{bbm}

\usepackage{relsize}
\usepackage{graphicx}

\newcommand{\newsection}{    
\setcounter{equation}{0}\section}
\def\appendix#1{\addtocounter{section}{1}\setcounter{equation}{0}
\renewcommand{\thesection}{\Alph{section}}
\section*{Appendix \thesection\protect\indent \parbox[t]{11.15cm}{#1}}
\addcontentsline{toc}{section}{Appendix \thesection\ \ \ #1}}

\def\bbe{{\bf{e}}}
\font\mybb=msbm10 at 11pt

\def\bb#1{\hbox{\mybb#1}}

\def\bZ {\bb{Z}}
\def\bR {\bb{R}}

\def\bS {\bb{S}}
\def\bCP{\bb{CP}}
\def\cL {{\cal{L}}}

\def\cD {{\cal{D}}}

\def\tn{{\tilde{\nabla}}}

\def\cL{{\cal L}}

\def\pl{{\parallel}}

\def\bbe{{\bf{e}}}
\font\mybb=msbm10 at 11pt

\def\bb#1{\hbox{\mybb#1}}

\def\bZ {\bb{Z}}
\def\bR {\bb{R}}

\def \la {\label}

\newcommand{\bea}{\begin{eqnarray}}
\newcommand{\eea}{\end{eqnarray}}

\begin{document}
\begin{titlepage}
\begin{center}
\vspace*{-1.0cm}
\hfill DMUS-MP-17-11
\\

\vspace{2.0cm} {\Large \bf All superalgebras for warped AdS$_2$ and  black hole  near horizon geometries} \\[.2cm]

\vspace{1.5cm}
 {\large U. Gran$^1$, J. Gutowski$^2$ and  G. Papadopoulos$^3$}

\vspace{0.5cm}

${}^1$ Department of Physics\\
Division for Theoretical Physics\\
Chalmers University of Technology\\
SE-412 96 G\"oteborg, Sweden\\

\vspace{0.5cm}
$^2$ Department of Mathematics \\
University of Surrey \\
Guildford, GU2 7XH, UK \\

\vspace{0.5cm}
${}^3$ Department of Mathematics\\
King's College London\\
Strand\\
London WC2R 2LS, UK\\

\end{center}

\vskip 3.5 cm
\begin{abstract}

We  identify all symmetry superalgebras $\mathfrak{g}$ of near horizon geometries of black holes with a
 Killing horizon, assuming the solution is smooth and that the spatial cross section of the
event horizon is compact without boundary.  This includes all  warped AdS$_2$ backgrounds with the most general allowed fluxes in  10- and 11-dimensional supergravities.  If the index of a particular Dirac operator vanishes,  we find that the even symmetry subalgebra decomposes as $\mathfrak{g}_0=\mathfrak{sl}(2,\bR)\oplus \mathfrak{t}_0$, where   $\mathfrak{t}_0/\mathfrak{c}$  is the Lie algebra of a group that acts transitively and effectively on spheres, and $\mathfrak{c}$ is the center of $\mathfrak{g}$. If  the Dirac operator index does not vanish, then the symmetry superalgebra is nilpotent with one even generator.
We also demonstrate that there are no near horizon geometries, and also therefore no warped AdS$_2$ backgrounds,  in 10- and 11-dimensions that preserve more than 16 supersymmetries.

\end{abstract}

\end{titlepage}



\section{Introduction}

A consequence of the horizon conjecture \cite{iibhor} is    that under certain conditions the near  horizon geometries of black holes with a Killing horizon  preserve
\bea
N=2 N_-+\mathrm{Index}\slashed D~,
\label{nindex}
\eea
  supersymmetries, where $\slashed D$ is a (twisted)  Dirac operator and $N_-\in \bZ_{\geq 0}$.  In addition   if $N_-\not=0$ and the near horizon geometries exhibit non-trivial fluxes,  their symmetry superalgebra contains  an $\mathfrak{sl}(2,\bR)$ subalgebra. This conjecture has been   demonstrated\footnote{The assumptions needed to prove these results will be explained below.} for  several theories in \cite{5hor}-\cite{4hor}.   These include the 10- and 11-dimensional supergravities as well as the ${\cal N}=2$  four- and ${\cal N}=1$ five-dimensional (gauged)  supergravities. The purpose of this paper is to identify all symmetry superalgebras of near  horizon geometries that satisfy the horizon conjecture.
  Warped AdS$_2$ backgrounds with the most general allowed fluxes are special cases of near horizon geometries.
  As a consequence, all symmetry superalgebras
of warped AdS$_2$ backgrounds in 10- and 11- as well as those in 4- and 5-dimensional theories will also be determined. These results together with those presented in
\cite{AdSsuperalgebras} complete the list\footnote{See \cite{charles} for the classification of symmetry superalgebras of warped AdS$_4$ backgrounds under different assumptions.} of all symmetry superalgebras\footnote{In \cite{AdSsuperalgebras} the symmetry superalgebras were referred to as Killing superalgebras.  We have changed the terminology as we use here the term Killing for the horizons.}
 of warped AdS$_n$ backgrounds for $n\geq 2$ with the most general allowed fluxes in 10- and 11-dimensional
supergravity theories.

One of our results is that if $\mathrm{Index}\, \slashed D=0$, then the even subalgebra $\mathfrak{g}_0$  of the symmetry superalgebra $\mathfrak{g}=\mathfrak{g}_0\oplus \mathfrak{g}_1$ of near horizon geometries decomposes as
$\mathfrak{g}_0=\mathfrak{sl}(2\, \bR)\oplus \mathfrak{t}_0$.  Moreover, $\mathfrak{t}_0/\mathfrak{c}$ is the Lie algebra of a group which acts transitively and effectively  on a sphere, where $\mathfrak{c}$ is the centre of $\mathfrak{g}$. These results allow
us to identify all possible such superalgebras which  have been tabulated in table 1.  These superalgebras are the same as either the left or the right copy
of the symmetry superalgebra of warped AdS$_3$ backgrounds found in \cite{AdSsuperalgebras}. Warped AdS$_2$ backgrounds exhibit the same symmetry superalgebras as near horizon geometries.
 However,   the geometry of the spatial horizon sections of generic near horizon geometries is different from that of the internal spaces of warped AdS$_2$ backgrounds.  In particular,
 the spatial horizon sections of generic near horizon geometries admit a  $\mathfrak{t}_0\oplus \mathfrak{so}(2)$ Lie algebra of isometries while the internal spaces of  AdS$_2$
 backgrounds admit a $\mathfrak{t}_0$ Lie algebra of isometries.

 The index of $\slashed D$ vanishes for all non-chiral theories.  For example it vanishes in 11-dimensional, (massive) IIA and 5-dimensional supergravities.  However it may not vanish
 in chiral theories like the IIB supergravity.  If $\mathrm{Index}\slashed D\not=0$ and  $N_-=0$, we demonstrate that the symmetry superalgebra of the backgrounds is nilpotent, $\mathfrak{g}_0$ is 1-dimensional and the non-vanishing anti-commutators are given in (\ref{nilanti}).  We also demonstrate that if $N_-\not=0$ and the theory has non-trivial fluxes, then $\mathrm{Index}\slashed D=0$ and thus the symmetry superalgebras
 are given as in table 1.

 Moreover, we show that for near horizon geometries that preserve more than half of supersymmetry,  the spatial horizon sections admit a transitive  group action of a group $G$
  with Lie algebra $\mathfrak{t}_0\oplus \mathfrak{so}(2)$. This follows from  an adaptation of the homogeneity theorem of \cite{homogen}.  Moreover    the action of the subgroup with Lie algebra  $\mathfrak{t}_0/\mathfrak{c}$ is effective.  Similarly  the internal spaces of warped AdS$_2$ backgrounds admit a transitive action of a group   $G$ with Lie algebra $\mathfrak{t}_0$ such that
  $\mathfrak{t}_0/\mathfrak{c}$ acts effectively.  We use these results together with the classification of  homogeneous spaces in  8 and 9 dimensions \cite{klaus} to demonstrate that there are no near horizon geometries and  warped AdS$_2$ backgrounds  in
10- and 11-dimensional supergravity theories that preserve more than 16 supersymmetries.

As has already been mentioned, the proof of the above results relies on the horizon conjecture which allows us to establish the results in a theory independent way.  In turn,  the assumptions required to establish  the horizon conjecture are  that the fields are  smooth and the black hole spatial horizon section is compact without boundary.
 The assumptions of the horizon conjecture are also needed  to exclude the possibility that warped AdS$_2$ backgrounds backgrounds can be rewritten as warped AdS$_n$, $n>2$,  backgrounds  \cite{strominger, desads}. If this possibility is not excluded,
symmetry superalgebras are different and in particular they do not decompose as $\mathfrak{g}_0=\mathfrak{sl}(2,\bR)\oplus \mathfrak{t}_0$.

To identify the symmetry superalgebra of near horizon geometries, we use the methodology developed in \cite{AdSsuperalgebras} to find the symmetry superalgebras
of warped AdS$_n$, $n>2$,  backgrounds. First, one can compute  part of the symmetry superalgebra  using the explicit  expression of the Killing spinors in terms of the light-cone coordinates
that arise in the description of  near horizon geometries. For this one utilizes   the geometric approach of defining  the brackets proposed in \cite{pktsuper, josesuper} via the use of Killing spinor bilinears and spinorial Lie derivatives
which is summarized in section 2. To identify the remaining brackets,  the closure of the symmetry superalgebra is imposed via the application of  super-Jacobi identities.  Closure of the symmetry superalgebras of supersymmetric backgrounds
 has been demonstrated in \cite{11jose, iibjose} for some theories and it is thought that it is valid  in general.
The advantage of imposing  the super-Jacobi identities  is that the computation of the remaining (anti-) commutators of the symmetry superalgebra can be done generically and it does not use details
of the geometry of the underlying backgrounds which may  not be available.

This paper is organized as follows.  In section 2, we summarize the geometry of supersymmetric near horizon geometries and review the construction of symmetry superalgebras.  In section 3, we construct the symmetry superalgebras of near horizon geometries and AdS$_2$ backgrounds preserving any number of supersymmetries.  In section 4,
we prove a no-go theorem for near horizon geometries and warped AdS$_2$ backgrounds that preserve more than half of supersymmetry, and in section 5 we give our conclusions.

\newsection{Horizons and  symmetry  }

\subsection{Near horizon geometries and warped AdS$_2$ backgrounds }

\subsubsection{Fields}

Consider a supergravity  theory with bosonic fields given by the metric $ds^2$ and some form field strengths which we shall denote collectively by $F$.
Adapting Gaussian null coordinates \cite{isen, gnull} near the Killing horizon of an extreme black hole, and after taking the near horizon limit,  the near horizon metric and form field strengths $F$ can be written \cite{smhor, mhor1} as
 \begin{eqnarray}
ds^2 &=& 2 {\bf{e}}^+ {\bf{e}}^- + \delta_{ij}\, {\bf{e}}^i {\bf{e}}^j=2 du (dr + r h - {1 \over 2} r^2 \Delta du)+ ds^2({\cal S})~,
\cr
F &=& {\bf{e}}^+ \wedge {\bf{e}}^- \wedge Y
+ r {\bf{e}}^+ \wedge M + X~,
\label{mhm}
\end{eqnarray}
where  we have introduced the frame
\begin{eqnarray}
{\bf{e}}^+ = du~,~~~{\bf{e}}^- = dr + r h_i{\bf e}^i - {1 \over 2} r^2 \Delta du~,~~~
{\bf{e}}^i = e^i{}_J dy^J~,
\label{nhbasis}
\end{eqnarray}
and
\begin{eqnarray}
ds^2({\cal S})=\delta_{ij} {\bf{e}}^i{\bf{e}}^j~,
\end{eqnarray}
is the metric of the spatial horizon section ${\cal S}$ given by $r=u=0$. The vector field $\partial_u$ is taken to be time-like and becomes null at the horizon hyper-surface $r=0$.
The dependence on the (light-cone) coordinates $r,u$ is given explicitly. The fields   $h=h_i {\bf{e}}^i$, $\Delta$, $ds^2({\cal S})$,  $Y$, $M$ and $X$ depend only on  the coordinates $y$
of ${\cal S}$.  If $F$ is a $k$-form field strength, then $Y$, $M$ and $X$ are $(k-2)$-, $(k-1)$- and $k$-forms on ${\cal S}$, respectively.  Furthermore,   $h$ and $\Delta$ are a 1-form and a 0-form
on ${\cal S}$, respectively.

The assumptions required for the validity of the horizon conjecture are that
\begin{itemize}
\item ${\cal S}$ is compact without boundary and

\item the fields
 $\Delta$,  $h$, $Y$, $M$  and $X$ are globally defined and smooth on ${\cal S}$.
 \end{itemize}
These are used in the proof of Lichnerowicz type  theorems which are necessary to establish the horizon conjecture, see \cite{iibhor}-\cite{4hor}.

The warped AdS$_2$ backgrounds with the most general allowed fluxes arise as a special case of near horizon geometries.  For this
one sets
\bea
\Delta=\ell^{-2} A^{-2}~,~~~h=-d\log A^2~,~~~M=0~,
\eea
where $A^2$ is a function of ${\cal S}$ and is identified with the warp factor, and $\ell$ is the radius of AdS$_2$.   $M$ is set to zero
so that the fluxes $F$ are invariant under the $\mathfrak{sl}(2,\bR)$ isometries of AdS$_2$. For more details, see \cite{adshor}.
 \subsubsection{Horizon conjecture and Killing spinors}

The solutions of the  Killing spinor equations (KSEs) of supergravity theories along the light-cone directions $r,u$ of a near horizon geometry given  in (\ref{mhm}) can be expressed as
\bea
\epsilon_1 = \phi_- + u \Gamma_+ \Theta_- \phi_- + ru \Gamma_- \Theta_+ \Gamma_+ \Theta_- \phi_-~,~~~
\epsilon_2 = \phi_+ + r \Gamma_- \Theta_+ \phi_+~,
\label{kkksp1x}
\eea
where $\phi_\pm$ are (commuting) spinors that depend only on the coordinates of ${\cal S}$, satisfy the light-cone
 projections $\Gamma_\pm\phi_\pm=0$  and are in the same representation of the spin group as that of the supersymmetry parameters of the underlying theory.  The spinors $\phi_\pm$ obey additional KSEs which can be thought of as the restriction of the KSEs of the theory
to the spatial horizon section ${\cal S}$.  These are theory dependent but their explicit form is not essential for the derivation
of the symmetry superalgebras of the near horizon geometries.  Moreover $\Theta_\pm$ are algebraic Clifford algebra operators along ${\cal S}$ which depend on the fields $\Delta, h, Y, M$ and $X$
of the theory  considered. The details of how $\Theta_\pm$  depend on the fields
is not again essential for the arguments that follow. $\Theta_\pm$ are independent of $\Gamma_\pm$ and so preserve the light-cone chirality of the spinors. The dependence of the Killing spinors in (\ref{kkksp1}) on the coordinates $u,r$   is explicit.

 It can be shown using  Hopf maximum principle and the KSEs on ${\cal S}$ that
 \bea
 \pl \phi_+\pl=\mathrm{const}~.
 \eea
 Therefore the length\footnote{The norm is taken with respect to an inner product $\langle\cdot, \cdot\rangle$ which is the real part of the Hermitian inner product for which gamma matrices along spatial directions are Hermitian
  while those along time-like directions are anti-hermitian.} of $\phi_+$ Killing spinors is constant.
 In addition, one can show that if $\phi_-\not=0$ solves the KSE on ${\cal S}$, i.e.~$N_-\not=0$, then
 \bea
 {{\phi}}_+ = \Gamma_+ \Theta_- \phi_-~,~~
 \eea
 also solves the KSEs on ${\cal S}$. Furthermore, one can demonstrate  using the Hopf maximum principle that for horizons with {\it  non-trivial
fluxes}  the kernel of $\Theta_-$ vanishes
\bea
\mathrm{Ker}\, \Theta_-=\{0\}~.
\label{vanker}
\eea
So, if $\phi_-$ is a Killing spinor then so is ${{\phi}}_+ = \Gamma_+ \Theta_- \phi_-\not=0$.
This is the origin of the doubling of supersymmetries for near horizon geometries with $N_-\not=0$.
  For the  derivation of (\ref{kkksp1x}) as well as all the remaining statements above see \cite{iibhor}-\cite{4hor}.

\subsubsection{$\mathfrak{sl}(2,\bR)$}

For any pair of Killing spinors $\epsilon_1$ and $\epsilon_2$,  one can define the spacetime 1-form bilinear as
\bea
K(\epsilon_1, \epsilon_2)=\langle (\Gamma_+-\Gamma_-) \epsilon_1, \Gamma_M \epsilon_2\rangle\, dx^M~.
\label{def1form}
\eea
It is expected that all these 1-forms give rise to spacetime Killing vectors that leave all the remaining fields of the theory
invariant.

Next consider the theories with $N_-\not=0$ and with non-trivial fluxes.  In this case there is a $\phi_-$ Killing spinor and which gives rise to
another Killing spinor given by ${{\phi}}_+ = \Gamma_+ \Theta_- \phi_-$.  Using these to construct the spacetime Killing spinors, one finds
\bea
\epsilon_1 = \phi_- + u \phi_+  + ru \Gamma_- \Theta_+ \phi_+ ~,~~~
\epsilon_2 = \phi_+ + r \Gamma_- \Theta_+ \phi_+~.
\label{kkksp1}
\eea
In particular set $K_{AB}=K(\epsilon_A, \epsilon_B)$, $A,B=1,2$,  for the Killing spinors $\epsilon_A$ in (\ref{kkksp1}). The requirement that $K_{AB}$ are Killing leads to the conditions
\bea
- \Delta\, \pl\phi_+\pl^2 +4  \pl\Theta_+ \phi_+\pl^2 =0~,~~~\langle \phi_+ , \Gamma_i \Theta_+ \phi_+ \rangle  =0~,
\label{deltai}
\eea
which in turn can be used to simplify the expression for $K_{AB}$ as
\bea
 K_{12}&=& ( 2r \langle\Gamma_+\phi_-, \Theta_+\phi_+\rangle+ r^2u \Delta ||\phi_+||^2) \,\bbe^+-2u \pl\phi_+\pl^2\, \bbe^-+ W_i \bbe^i~,
 \cr
 K_{22}&=& r^2 \Delta\pl\phi_+\pl^2 \,\bbe^+-2 \pl\phi_+\pl^2 \bbe^-~,
 \cr
 K_{11}&=&(2\pl\phi_-\pl^2+4r u \langle\Gamma_+\phi_-, \Theta_+\phi_+\rangle+ r^2 u^2 \Delta \pl\phi_+\pl^2) \bbe^+
 \cr
 && \qquad\qquad \qquad\qquad -2u^2 \pl\phi_+\pl^2 \bbe^-+2u W_i \bbe^i~,
 \label{1fbiln2}
 \eea
where we have set
\bea
\label{extraiso}
W_i =  \langle \Gamma_+ \phi_- , \Gamma_i \phi_+ \rangle~.
\eea
The relations (\ref{deltai}) also follow from the KSEs on ${\cal S}$ and the constancy of  $\pl\phi_+\pl^2$.
Furthermore, the Killing condition on $K_{AB}$   implies the relations
\bea
i_Wd\Delta=0~,~~~\tilde\nabla_{(i} W_{j)}=0~,
\eea
and
\bea
&&-2 \pl\phi_+\pl^2-h_i W^i+2 \langle\Gamma_+\phi_-, \Theta_+\phi_+\rangle=0~,~~~i_W (dh)+2 d \langle\Gamma_+\phi_-, \Theta_+\phi_+\rangle=0~,
\cr
&& 2 \langle\Gamma_+\phi_-, \Theta_+\phi_+\rangle-\Delta \pl\phi_-\pl^2=0~,~~~
W+ \pl\phi_-\pl^2 h+d \pl\phi_-\pl^2=0~,
\la{concon}
\eea
where $\tilde \nabla$ is the Levi-Civita connection on ${\cal S}$.
Using the above conditions, one can also show that
\bea
\tilde {\cal L}_Wh=0~,~~\tilde {\cal L}_W\pl\phi_-\pl^2=0~,
\eea
where $\tilde {\cal L}$ is the Lie derivative\footnote{To distinguish operations along ${\cal S}$ from those on the spacetime, we denote the former using a tilde.} in ${\cal S}$.
The invariance of all of the fields of the theory under $K_{AB}$ also implies that $W$  leaves invariant
all the components of these fields that have support on the spatial horizon section. For example,
 the invariance of the form flux $F$  under $K_{AB}$ implies
 that the components $Y$, $M$  and $X$ of $F$
 are invariant\footnote{If the fluxes are twisted like in IIB, then the invariance is up to gauge transformations.} under $W$,  ${\tilde {\cal L}}_WY={\tilde {\cal L}}_WM={\tilde {\cal L}}_WX=0$.

Using the relations (\ref{concon}), the associated vector fields to $K_{AB}$ are
\bea
K_{12}&=&-2u \pl\phi_+\pl^2 \partial_u+ 2r \pl\phi_+\pl^2 \partial_r+ W^i \tilde \partial_i~,~~~
K_{22}=-2 \pl\phi_+\pl^2 \partial_u~,
\cr
K_{11}&=&-2u^2 \pl\phi_+\pl^2 \partial_u +(2 \pl\phi_-\pl^2+ 4ru \pl\phi_+\pl^2)\partial_r+ 2u W^i \tilde \partial_i~.
\label{kvn2}
\eea
One can easily compute the Lie algebra of these vector fields  to find
\bea
[K_{AB}, K_{A'B'}]&=&\pl \phi_+\pl^2 (\epsilon_{AA'} K_{BB'}+ \epsilon_{BB'} K_{AA'}
\cr
&&\qquad \qquad+\epsilon_{BA'} K_{AB'}+\epsilon_{AB'} K_{BA'})~,
\eea
where $\epsilon_{AB}=-\epsilon_{BA}$ and $\epsilon_{12}=1$. This  is isomorphic to  $\mathfrak{sl}(2,\bR)$.  For the proof  on how the  $\mathfrak{sl}(2,\bR)$ symmetry
arises in the context of horizons as a consequence of the horizon conjecture see \cite{iibhor}-\cite{4hor}.


\subsection{Symmetry superalgebras}

The symmetry superalgebras of  supersymmetric backgrounds are constructed as follows. First one
introduces for each Killing spinor $\epsilon_{\mathrm m}$ of the background an odd generator $Q_{\mathrm m}=Q_{\epsilon_{\mathrm m}}$  and for
each 1-form bilinear $K_{\mathrm {mn}}=K(\epsilon_{\mathrm m}, \epsilon_{\mathrm n})$ an even generator $V_{\mathrm {mn}}=V_{K_{\mathrm {mn}}} $ for the superalgebra, respectively.
The symmetry superalgebra of the
background is spanned as $\mathfrak{g}=\mathfrak{g}_0\oplus \mathfrak{g}_1=\bR\langle V_{\mathrm{mn}},  Q_{\mathrm p}\rangle$, where $\mathfrak{g}_0=\bR\langle V_{\mathrm{mn}}\rangle$ is the even subalgebra and $\mathfrak{g}_1=\bR\langle Q_{\mathrm p}\rangle$ is the odd subspace.
The (anti-) commutators
of $\mathfrak{g}$ are defined \cite{pktsuper, josesuper} as
\bea
\{Q_{\mathrm m}, Q_{\mathrm n}\}= V_{\mathrm{mn}}~,~~~[V_{\mathrm{mn}}, Q_{\mathrm p}] = Q_{{\cal L}_{K_{\mathrm{mn}}}\epsilon_{\mathrm p}}~,~~~[V_{\mathrm{mn}}, V_{\mathrm{m}'\mathrm{n}'}]=
V_{[K_{\mathrm{m} \mathrm{n}}, K_{\mathrm{m}'\mathrm{n}'}]}
\eea
where ${\cal L}_{K_{\mathrm{mn}}}$ is the spinorial Lie derivative\footnote{The spinorial Lie derivative of the spinor $\epsilon$ along the Killing vector field $K$ is ${\cal L}_K\epsilon=\nabla_K\epsilon+{1\over8} {\slashed{dK}}\epsilon$.} along the vector field $K_{\mathrm{mn}}$
and the commutator in the right-hand-side in the last equation is that of two vector fields.  The closure of this superalgebra for supersymmetric backgrounds of some supergravity
theories which include 10- and 11-dimensional supergravities has been demonstrated in \cite{11jose, iibjose}.

\section{Superalgebras of horizons and AdS$_2$ backgrounds}

\subsection{The superalgebra of  $N=2$ horizons with $\mathrm{Index}\,\slashed D=0$}

As  $\mathrm{Index}\slashed D=0$, the Killing spinors  of near horizon geometries preserving two supersymmetries, $N=2$, are given as in (\ref{kkksp1}) and the 1-form bilinears as in (\ref{1fbiln2}). The symmetry superalgebra  is spanned as
$\mathfrak{g}=\bR\langle V_{AB}, Q_C\rangle$.  The anti-commutators $\{Q_A, Q_B\}$ can be extracted from the
1-form bilinears (\ref{1fbiln2}). To find the commutators of even with odd generators, we have to compute the
spinorial derivative of the Killing spinors (\ref{kkksp1}) with respect to the Killing vectors (\ref{kvn2}).
After some computation using the relations (\ref{concon}), one finds that the  a priori non-vanishing spinorial Lie derivatives are
\bea
&&{\cal L}_{K_{12}} \epsilon_1=-\pl\phi_+\pl^2 \epsilon_1+ \tilde{{\cal L}}_W \epsilon_1~,~~~{\cal L}_{K_{12}} \epsilon_2= \pl\phi_+\pl^2 \epsilon_2+ \tilde{{\cal L}}_W \epsilon_2~,
\cr
&&{\cal L}_{K_{22}} \epsilon_1= -2\pl\phi_+\pl^2 \epsilon_2~,
\cr
&&{\cal L}_{K_{11}} \epsilon_1= u(-2 \pl\phi_+\pl^2 \phi_-+ 2 \pl\phi_-\pl^2 \Gamma_-\Theta_+\phi_+-
\slashed{W} \Gamma_-\phi_+)+ 2u \tilde {{\cal L}}_W \epsilon_1~,
\cr
&& {\cal L}_{K_{11}} \epsilon_2= 2\pl\phi_+\pl^2 u \epsilon_2+ 2 \pl\phi_-\pl^2 \Gamma_-\Theta_+\phi_+
-\slashed{W} \Gamma_-\phi_++ 2u \tilde {{\cal L}}_W \epsilon_2~.
\eea
 Closure of the Killing superalgebra requires first that these spinorial Lie derivatives
must be expressed in terms of the original Killing spinors $\epsilon_A$ and second that the super-Jacobi identities of the symmetry superalgebra
must be satisfied. For the spinorial Lie derivatives above to satisfy these criteria additional conditions  must be imposed.

The  super-Jacobi identity of the symmetry superalgebra for three $Q_1$ generators is satisfied provided that
${\cal L}_{K_{11}} \epsilon_1=0$.  This in particular gives  the conditions
\bea
\tilde {{\cal L}}_W \phi_+=0~,~~~\tilde {{\cal L}}_W\Theta_+ \phi_+=0~,
\label{oncon}
\eea
and
\bea
2 \tilde {{\cal L}}_W \phi_--2 \pl\phi_+\pl^2 \phi_-+ 2 \pl\phi_-\pl^2 \Gamma_-\Theta_+\phi_+-
\slashed{W} \Gamma_-\phi_+=0~.
\la{seccon}
\eea
Typically, the second condition in (\ref{oncon}) follows from the
first and the invariance of the fields on ${\cal S}$ under the action of $W$.

Using (\ref{seccon}) as well as (\ref{deltai}) and (\ref{concon}), one can show that
$\pl \tilde {{\cal L}}_W \phi_-\pl^2=0$
and
so
\bea
 \tilde {{\cal L}}_W \phi_-=0~.
\label{invphim}
\eea
Because of this,  (\ref{seccon}) gives
\bea
-2 \pl\phi_+\pl^2 \phi_-+ 2 \pl\phi_-\pl^2 \Gamma_-\Theta_+\phi_+-
\slashed{W} \Gamma_-\phi_+=0~,
\label{phipphim}
\eea
and (\ref{oncon}) together with (\ref{invphim}) lead to
\bea
\tilde {{\cal L}}_W\epsilon_1=\tilde {{\cal L}}_W\epsilon_2=0~.
\eea
The non-vanishing spinorial Lie derivatives can now be re-arranged as
\bea
&&{\cal L}_{K_{12}} \epsilon_1=-\pl\phi_+\pl^2 \epsilon_1~,~~~{\cal L}_{K_{12}} \epsilon_2= \pl\phi_+\pl^2 \epsilon_2~,
\cr
&&{\cal L}_{K_{22}} \epsilon_1= -2\pl\phi_+\pl^2 \epsilon_2~,~~~ {\cal L}_{K_{11}} \epsilon_2= 2\pl\phi_+\pl^2  \epsilon_1~.
\label{spilien2}
\eea
Note that to establish the above spinorial Lie derivatives, we did not make use of the details of the geometry of the underlying spacetime.
Furthermore notice that (\ref{phipphim}) can be seen as the inverse transformation to $\phi_+=\Gamma_+\Theta_-\phi_-$ as it expresses
implicitly $\phi_-$ in terms of $\phi_+$.

The spinorial Lie derivatives (\ref{spilien2}) specify all the commutators between the even and odd generators of the symmetry superalgebra. Collecting all non-vanishing (anti-) commutators together we have
\bea
&&\{Q_A, Q_B\}=V_{AB}~,~~~[V_{AB},  Q_C]=- \epsilon_{CA} Q_B-\epsilon_{CB} Q_A~,
\cr
&& [V_{AB}, V_{A'B'}]=\epsilon_{AA'} V_{BB'}+ \epsilon_{BB'} V_{AA'}
+\epsilon_{BA'} V_{AB'}+\epsilon_{AB'} V_{BA'}~,
\label{osp1n2}
\eea
where we have set  $\pl\phi_+\pl^2=1$.
 Therefore the symmetry superalgebra $\mathfrak{g}=\bR\langle V_{AB}, Q_C\rangle$ of near horizon geometries preserving strictly two supersymmetries is isomorphic to  $\mathfrak{osp}(1|2)$.
 Note that there is no superalgebra generator corresponding to the Killing vector field $W$.  Instead the generators $V_{AB}$ are associated with vector fields which have components
 both along the lightcone coordinates and along directions tangent to ${\cal S}$. This is because $W$ is not associated with a spacetime 1-form
 spinor bilinear.

\subsection{Superalgebra of $N=2$  AdS$_2$ backgrounds with $\mathrm{Index}\,\slashed D=0$}

Warped AdS$_2$ backgrounds with the most general allowed fluxes are special cases of near horizon geometries.  These arise whenever $W=0$.  Substituting this into
the last equation in (\ref{concon}) gives
\bea
\Delta=\ell^{-2} \pl\phi_-\pl^{-2}~,~~~ h=-d\log\pl\phi_-\pl^2~,~~~~\pl\phi_+\pl^2={1\over 2\ell^2}~.
\eea
Therefore spacetime  is  a warped AdS$_2$ space, $AdS_2\times_w {\cal S}$, with warp factor $A^2=\pl\phi_-\pl^2$ \cite{desads}. The symmetry superalgebra of warped AdS$_2$ backgrounds preserving
two supersymmetries is again $\mathfrak{osp}(1|2)$.  The only difference between AdS$_2$ backgrounds and generic near horizon geometries is that in the former case the orbits of $\mathfrak{sl}(2,\bR)$ subalgebra are 2-dimensional and so all three associated vector fields (\ref{kvn2}) are tangent to the AdS$_2$ subspace,  while in the latter case the orbits of
$\mathfrak{sl}(2,\bR)$  can be 3-dimensional.

\subsection{Horizons and AdS$_2$ backgrounds  with $N>2$ and $\mathrm{Index}\,\slashed D=0$}

\subsubsection{Killing spinors and spinorial Lie derivatives}
Suppose now that a near horizon geometry of an extreme  black hole with a Killing horizon  admits $N=2k$ supersymmetries and exhibits non-trivial fluxes. The horizon conjecture implies that the Killing spinors
can be chosen as
\begin{eqnarray}
\epsilon^r_{1} &=& \phi^s_- + u {{\phi}}^s_+ + ru \Gamma_- \Theta_+ {{\phi}}^s_+~,
\nonumber \\
\epsilon^r_{2} &=& \phi^s_+ + r \Gamma_- \Theta_+ \phi^s_+~,
\label{mkkspx}
\end{eqnarray}
where ${{\phi}}^s_+ = \Gamma_+ \Theta_- \phi^s_-$ and  $\phi^r_-$, $s=1,\dots, k$ are linearly independent Killing spinors on the spatial horizon section ${\cal S}$.
In addition the horizon conjecture implies that
\bea
\langle \phi_+^r, \phi_+^s\rangle=\mathrm{const}~,
\label{const2}
\eea
and that   the matrix $(\langle \phi_+^r, \phi_+^s\rangle)$ is non-degenerate because of (\ref{vanker}).  In what follows without loss of generality we shall take $(\langle \phi_+^r, \phi_+^s\rangle)$
to be proportional to the identity matrix though we shall leave $\langle \phi_+^r, \phi_+^s\rangle$ in the formulae below as this illustrates better the origin and the meaning  of some of the terms.

To proceed, one uses (\ref{def1form}) to  define the 1-form Killing spinor bi-linears $K_{AB}^{rs}=K(\epsilon_A^r, \epsilon_B^s)$.   The Killing condition on  $K_{AB}^{rs}$ gives
 \bea
4 \langle \Theta_+ \phi_+^r,
\Theta_+ \phi_+^s\rangle= \Delta \langle \phi_+^r, \phi_+^s\rangle~,~~~\langle \phi_+^{(r}, \Gamma_i\Theta_+\phi_+^{s)}\rangle=0~,
\eea
which can be used to simplify the expression for the bilinears as
\bea
&&K^{rs}_{12}= \big(2r \langle \Gamma_+\phi_-^r, \Theta_+\phi_+^s\rangle+  r^2 u \Delta \langle \phi_+^r, \phi_+^s\rangle\big)\, \bbe^+- 2u \langle \phi_+^r, \phi_+^s\rangle\, \bbe^-+  W^{rs}~,
\cr
&&K^{rs}_{22}= r^2 \Delta \langle \phi_+^r, \phi_+^s\rangle\, \bbe^+- 2 \langle \phi_+^r, \phi_+^s\rangle \bbe^-~,
\cr
&&K^{rs}_{11}= \big(2 \langle \phi_-^r, \phi_-^s\rangle  +4ru   \langle \Gamma_+\phi_-^{(r}, \Theta_+\phi_+^{s)}\rangle+  r^2 u^2 \Delta \langle \phi_+^r, \phi_+^s\rangle\big)\, \bbe^+
\cr
&&\qquad\qquad - 2u^2 \langle \phi_+^r, \phi_+^s\rangle \bbe^-+ 2u W^{(rs)}~,
\label{krsv}
\eea
where
\bea
W^{rs}=\langle \Gamma_+ \phi_-^r, \Gamma_i \phi_+^s\rangle\, \bbe^i~.
\eea
The Killing condition in addition requires
\bea
i_{W^{rs}}d\Delta=0~,~~~ {\tilde \nabla}_{(i} W_{j)}^{rs}=0~,
\label{ininin}
\eea
and
\bea
&&-2 \langle \phi_+^r, \phi_+^s\rangle-i_h W^{rs}+2 \langle\Gamma_+\phi^r_-, \Theta_+\phi^s_+\rangle=0~,~~~i_{W^{rs}} (dh)+2 d \langle\Gamma_+\phi^r_-, \Theta_+\phi^s_+\rangle=0~,
\cr
&& 2 \langle\Gamma_+\phi^{(r}_-, \Theta_+\phi^{s)}_+\rangle-\Delta \langle \phi_-^r, \phi_-^s\rangle=0~,~~~
W^{(rs)}+ \langle \phi_-^r, \phi_-^s\rangle  h+d \langle \phi_-^r, \phi_-^s\rangle=0~.
\la{conconrs}
\eea
Therefore all vector fields $W^{rs}$ are Killing on ${\cal S}$. It follows from (\ref{conconrs}) that
\bea
\tilde {\cal L}_{W^{rs}} h=0~.~~~
\eea
Furthermore in all  theories for which  the invariance of the remaining fields $F$  of the theory under $K_{AB}^{rs}$ has been investigated, it has been found to imply
that $Y,M$ and $X$ are invariant under $W^{rs}$ on ${\cal S}$.

To find the commutators of even and odd generators, we must compute the
spinorial Lie derivatives of the Killing spinors (\ref{mkkspx}) with respect to the
1-form bilinears (\ref{krsv}).   After some computation, one finds that the a priori
non-vanishing spinorial Lie derivatives are
\bea
{\cal L }_{K^{rs}_{12}}\epsilon_1^t&= &- \langle \phi_+^r, \phi_+^s\rangle \epsilon_1^t+ {\tilde {\cal L}}_{W^{rs}} \epsilon^t_1~,~~~
{\cal L}_{K^{rs}_{12}}\epsilon_2^t=  \langle \phi_+^r, \phi_+^s\rangle \epsilon_2^t+ {\tilde {\cal L}}_{W^{rs}} \epsilon^t_2~,
\cr
{\cal L }_{K^{rs}_{22}}\epsilon_1^t&= &-2 \langle \phi_+^r, \phi_+^s\rangle \epsilon^t_2~,
\cr
{\cal L}_{K^{rs}_{11}}\epsilon_1^t&= &u \big(-2 \langle \phi_+^r, \phi_+^s\rangle \phi_-^t
+ 2 \langle \phi_-^r, \phi_-^s\rangle \Gamma_- \Theta_+ \phi_+^t-{ \slashed W}^{(rs)}\Gamma_-\phi_+^t\big)+ 2u {\tilde{\cal L}}_{W^{(rs)}} \epsilon^t_1~,
\cr
{\cal L}_{K^{rs}_{11}}\epsilon_2^t&= & 2u  \langle \phi_+^r, \phi_+^s\rangle \epsilon_2^t
+2 \langle \phi_-^r, \phi_-^s\rangle \Gamma_-\Theta_+\phi_+^t- {\slashed W}^{(rs)}\Gamma_-\phi_+^t
+2u {\tilde{\cal L}}_{W^{(rs)}}
 \epsilon^t_2~.
\eea
Closure of the Killing superalgebra requires that ${\cal L}_{K^{rs}_{11}}\epsilon_1^t$ should
be expressed in terms of the original Killing spinors $\epsilon_A^r$.  However, this is not possible
as its dependence  on the lightcone coordinates $u,r$ is in conflict
with that of the Killing spinors.  This is unless all the coefficients vanish.  Thus we have that
\bea
&&{\tilde{\cal L}}_{W^{(rs)}} \phi_+^t= {\tilde{\cal L}}_{W^{(rs)}}\Theta_+ \phi_+^t =0~,~~~
\cr
&&2 {\tilde{\cal L}}_{W^{(rs)}} \phi_-^t= 2 \langle \phi_+^r, \phi_+^s\rangle \phi_-^t
 -2 \langle \phi_-^r, \phi_-^s\rangle \Gamma_- \Theta_+ \phi_+^t+{ \slashed W}^{(rs)}\Gamma_-\phi_+^t~.
 \label{cwphi1}
 \eea
Moreover as  ${{\phi}}^r_+ = \Gamma_+ \Theta_- \phi^r_-$, we  have
\bea
{\tilde{\cal L}}_{W^{(rs)}} \phi_+^t=0\Longrightarrow \Theta_-{\tilde{\cal L}}_{W^{(rs)}} \phi_-^t=0\Longrightarrow {\tilde{\cal L}}_{W^{(rs)}} \phi_-^t=0~,
\eea
where we have used that $h,\Delta$ and the fluxes $Y,M, X$ are invariant under $W^{(rs)}$, and that for near horizon geometries with non-trivial fluxes $\mathrm{Ker}\, \Theta_-=\{0\}$.
Hence (\ref{mkkspx}) and  (\ref{cwphi1}) imply that
\bea
{\tilde{\cal L}}_{W^{(rs)}} \epsilon_A^t=0~,~~~2 \langle \phi_+^r, \phi_+^s\rangle \phi_-^t
 -2 \langle \phi_-^r, \phi_-^s\rangle \Gamma_- \Theta_+ \phi_+^t+{ \slashed W}^{(rs)}\Gamma_-\phi_+^t=0~.
 \label{cwphi2}
\eea
Then the non-vanishing spinorial Lie derivatives can be re-arranged as
\bea
{\cal L }_{K^{rs}_{12}}\epsilon_1^t&= &- \langle \phi_+^r, \phi_+^s\rangle \epsilon_1^t+ {\tilde {\cal L}}_{Z^{rs}} \epsilon^t_1~,
\cr
{\cal L}_{K^{rs}_{12}}\epsilon_2^t&= & \langle \phi_+^r, \phi_+^s\rangle \epsilon_2^t+ {\tilde {\cal L}}_{Z^{rs}} \epsilon^t_2~,
\cr
{\cal L }_{K^{rs}_{22}}\epsilon_1^t&= &-2 \langle \phi_+^r, \phi_+^s\rangle \epsilon^t_2~,
\cr
{\cal L}_{K^{rs}_{11}}\epsilon_2^t&= & 2  \langle \phi_+^r, \phi_+^s\rangle \epsilon_1^t~,
\label{splienn}
\eea
where we have set $Z^{rs}=W^{[rs]}$.

Before we proceed to describe the superalgebra observe that the vector fields
$W^{rs}= (W^{rs})^i\tilde \partial_i$ are both isometries of the spatial horizon section
and also of the spacetime.

\subsubsection{$\mathfrak{g}_0=\mathfrak{sl}(2,\bR)\oplus \mathfrak{t}_0$}

To demonstrate that the even subalgebra decomposes as indicated above, we write
\bea
 K_{AB}^{(rs)}= \langle \phi_+^r, \phi_+^s\rangle K_{AB}+ {\mathring K}_{AB}^{(rs)}~,
 \eea
 where ${\mathring K}^{(rs)}$ is traceless with respect to the non-degenerate positive definite matrix $(\langle \phi_+^r, \phi_+^s\rangle)$.  Observe that the Killing vectors $K_{AB}$ have components both along the light-cone and spatial horizon
 section directions which generically  cannot be separated.

 We shall now demonstrate that
 \bea
 {\mathring K}_{AB}^{(rs)}=0~.
 \label{vancond}
 \eea
This is best done in the language of symmetry superalgebras. For this  introduce odd generators $Q^r_A$ for each spacetime Killing spinor $\epsilon_A^r$
 and even generators $V_{AB}^{rs}$ for each 1-form bilinear $K^{rs}_{AB}$. Writing
 \bea
 V_{AB}^{rs}= V^{(rs)}_{AB}+ \epsilon_{AB} \tilde V^{rs}=  \langle \phi_+^r, \phi_+^s\rangle V_{AB}+ {\mathring V}_{AB}^{(rs)}+\epsilon_{AB} \tilde V^{rs}~,
 \eea
 where $ V_{AB}^{rs}=\epsilon_{AB}\tilde V^{rs}$,   we have the (anti)-commutators
 \bea
 \{Q_A^r, Q_B^s\}=\langle \phi_+^r, \phi_+^s\rangle V_{AB}+ {\mathring V}_{AB}^{(rs)}+\epsilon_{AB} \tilde V^{rs}~,
 \eea
 and
 \bea
 [V_{AB}, Q_C^r]&=& -\epsilon_{CA} Q_B^r- \epsilon_{CB} Q_A^r~,
 \cr
 [{\mathring V}_{AB}^{(rs)}, Q_C^t]&=&0~.
 \label{vqqq}
 \eea
 The last commutator above follows from the spinorial Lie derivatives in (\ref{splienn}).
As  the ${\mathring V}_{AB}^{(rs)}$ commute with the $Q$'s, they are all central in the symmetry superalgebra $\mathfrak{g}$.

To continue observe that a super-Jacobi identity implies that
\bea
[V_{AB}^{rs}, V{}_{A'B'}^{r's'}]= \{ Q_{A'}^{r'}, [V_{AB}^{rs}, Q_{B'}^{s'}]\}+ \{Q_{B'}^{s'}, [V_{AB}^{rs}, Q_{A'}^{r'}]\}~.
\label{supjac}
\eea
Next set  $V_{AB}^{rs}= \langle \phi_+^r, \phi_+^s\rangle V_{AB}$  and $V{}_{A'B'}^{r's'}=  {\mathring V}_{A'B'}^{(r's')}$ in
(\ref{supjac}).  This gives
\bea
[V_{AB}, {\mathring V}_{A'B'}^{(r's')}]=\epsilon_{AB'} {\mathring V}_{B A'}^{(r's')}+\epsilon_{BB'} {\mathring V}_{AA'}^{(r's')}+ \epsilon_{AA'} {\mathring V}_{BB'}^{(r's')}+
\epsilon_{BA'} {\mathring V}_{AB'}^{(r's')}~.
\eea
As ${\mathring V}_{A'B'}^{(r's')}$ are central, they must commute with $V_{AB}$ and so  this commutator must vanish.  This is the case provided that
 ${\mathring V}_{A'B'}^{(r's')}=0$, which in turn implies that ${\mathring K}_{A'B'}^{(r's')}=0$.  This establishes (\ref{vancond}) and
 so
 \bea
 K^{rs}_{AB}= \langle \phi_+^r, \phi_+^s\rangle K_{AB}+ \epsilon_{AB} Z^{rs}~,~~~W^{rs}=\langle \phi_+^r, \phi_+^s\rangle \check W+ Z^{rs}~.
 \eea
 It remains to show that $[ K_{AB}, Z^{rs}]=0$.
To demonstrate this,  we use again (\ref{supjac}) with $V_{AB}^{rs}= \langle \phi_+^r, \phi_+^s\rangle V_{AB}$ and
$V{}_{A'B'}^{r's'}=\epsilon_{A'B'} \tilde V^{r's'}$. After applying (\ref{vqqq}),  a brief computation reveals that $[V_{AB},  \tilde V^{r's'}]=0$
which establishes the result.  In particular, this implies that
\bea
[Z^{rs}, \check W]=0~.
\label{commut}
\eea
This will be used later in the identification of the Lie algebras of groups that act transitively on the spatial horizon section.

\subsubsection{The commutator $[\tilde V^{rs}, Q^t_A]$}

To determine the symmetry superalgebra of near horizon geometries, it remains to find the  commutator $[\tilde V^{rs}, Q^t_A]$.  This
 can be done via  an analysis similar to that presented  in \cite{AdSsuperalgebras}
for warped AdS$_3$ backgrounds. Here we summarize some of the key points as there are some differences with the analysis presented for AdS$_3$ backgrounds.
The computation for the $N=4$ case is identical to that presented for AdS$_3$ backgrounds.  To identify the superalgebra for  the $N>4$ cases, one has to determine
$\tilde{\cal L}_{Z^{rs}} \epsilon_A^p$ for $p\not=r,s$.  First observe that $\tilde{\cal L}_{Z^{rs}}$ commutes with the lightcone coordinates and
so preserves the type of Killing spinors.  Thus
\bea
\tilde{\cal L}_{Z^{rs}} \epsilon_A^p=\alpha^{rsp}{}_q \epsilon_A^q~,~~~p\not= r,s~.
\eea
The constants $\alpha$ can be different for $\epsilon_1^p$ and $\epsilon_2^p$ Killing spinors.  However this is not the case as a consequence of the super-Jacobi identities
involving $V_{AB}$, $\tilde V^{rs}$ and $Q_A^p$.   In addition the super-Jacobi identities for $Q^r_A$, $Q^s_B$ and $Q^p_A$ for $B\not= A$ imply that
\bea
\alpha^{rsp}{}_q=\alpha^{prs}{}_q~,
\eea
which together with $\alpha^{rsp}{}_q=-\alpha^{srp}{}_q$ give $\alpha^{rsp}{}_q=\alpha^{[rsp]}{}_q$. Furthermore,  the spinorial Lie derivative identity
\bea
\tilde{\cal L}_{Z^{rs}}\langle \epsilon_2^p, \epsilon_2^q\rangle=\langle \tilde{\cal L}_{Z^{rs}}\epsilon_2^p, \epsilon_2^q\rangle+\langle \epsilon_2^p, \tilde{\cal L}_{Z^{rs}}\epsilon_2^q\rangle~,
\eea
together with
\bea
\tilde{\cal L}_{Z^{rs}}\langle \epsilon_2^p, \epsilon_2^q\rangle=\tilde{\cal L}_{Z^{rs}}(\langle \phi_+^p, \phi_+^q\rangle+{1\over2} r^2 \Delta \langle \phi_+^p, \phi_+^q\rangle)=0~,
\eea
which follows from (\ref{ininin}),  the invariance of $\Delta$  and (\ref{const2}),  give that $\alpha^{rspq}=\alpha^{rs[pq]}$, where the indices are raised with $\langle \phi_+^p, \phi_+^q\rangle$.  Putting all the properties of $\alpha$ together we find that
\bea
\alpha^{rspq}=\alpha^{[rspq]}~.
\eea
So $\alpha$ is an 4-form.  This conclusion is  the starting point of the analysis for the identification of the symmetry superalgebras of AdS$_3$ backgrounds.
The results are summarized in the section below.

 \subsubsection{The symmetry superalgebras of horizons with $\mathrm{Index}\,\slashed D=0$}

 Collecting all the results that we have established so far, the non-vanishing (anti-) commutators of the symmetry superalgebras of near horizon geometries
 are
 \bea
 &&\{Q_A^r, Q_B^s\}= \delta^{rs} V_{AB}+\epsilon_{AB} \tilde V^{rs}~,~~~[V_{AB}, Q_C^r]= -\epsilon_{CA} Q_B^r- \epsilon_{CB} Q_A^r~,
 \cr
&& [\tilde V^{rs}, Q^t_A]=- (\delta^{tr}  Q^s_A- \delta^{ts} Q^r_A- \alpha^{rst}{}_\ell Q^\ell_A)~,
 \cr
&& [V_{AB}, V_{A'B'}]= \epsilon_{AA'} V_{BB'}+ \epsilon_{BB'} V_{AA'} +\epsilon_{BA'} V_{AB'}+\epsilon_{AB'} V_{BA'}~,
\cr
&&[\tilde V^{rs}, \tilde V^{r's'}]=\delta^{rs'} \tilde V^{sr'}-\delta^{ss'} \tilde V^{rr'}-\delta^{rr'} \tilde V^{ss'}+ \delta^{sr'} \tilde V^{rs'}
\cr
&&\qquad\qquad\qquad - \alpha^{rss'}{}_t \tilde V^{tr'}+
\alpha^{rsr'}{}_t \tilde V^{ts'}~,
\label{finsup}
\eea
where we have set without loss of generality $ \langle \phi_+^r, \phi_+^s\rangle=\delta^{rs}$ to simplify the expressions.

The structure constants that remain to be determined are those that appear in the commutators $[\tilde V^{rs}, Q^t_A]$.  It turns out that $\alpha_{rst\ell}$ is a 4-form, where
the indices are raised and lowered with $\delta_{rs}$. The symmetry superalgebra $\mathfrak{g}$ defined in (\ref{finsup}) may have central generators.  Writing
\bea
[\tilde V^{rs}, Q^t_A]=D(V^{rs})^t{}_\ell\, Q^\ell_A~,
\eea
the centre $\mathfrak{c}$ is spanned by those elements  $a_{rs} \tilde V^{rs}\in \mathfrak{t}_0$ which commute with all the $Q$'s, i.e.  $a_{rs} D(\tilde V^{rs})^t{}_\ell=0$.
It has been shown in the context of AdS$_3$ backgrounds \cite{AdSsuperalgebras} that $\mathfrak{t}_0/\mathfrak{c}$ can be identified with the Lie algebra of a  group that acts transitively on a sphere with
the $D$ representation.  Moreover $\alpha$ is invariant under $D$. This together with the (anti-) commutators collected in (\ref{finsup}) allow to
identify all possible symmetry superalgebras of near horizon geometries that exhibit non-trivial fluxes and  $\mathrm{Index}\slashed D=0$.  The results have been tabulated in table 1. The same
symmetry superalgebras occur in warped AdS$_2$ backgrounds with the most general allowed fluxes, i.e.~the fluxes that are invariant under the $\mathfrak{sl}(2,\bR)$
isometries of AdS$_2$ subspace.

\begin{table}[h]
\begin{center}
\vskip 0.3cm
  \caption{Symmetry Superalgebras of near horizon geometries and warped AdS$_2$ backgrounds}
 \vskip 0.3cm
 \begin{tabular}{|c|c|c|}
  \hline
  $N$ & $\mathfrak{g}/\mathfrak{c}$& $\mathfrak{t}_0/\mathfrak{c}$
  \\ \hline
  $2n$  &$\mathfrak{osp}( n\vert 2)$ &$\mathfrak{so}( n)$
  \\ \hline
 $ 4n, n>1 $ & $\mathfrak{sl}(n\vert 2)$ & $\mathfrak{u}(n)$
  \\ \hline
 $8n, n>1$  &  $ \mathfrak{osp}^*(4\vert 2n)$ &  $ \mathfrak{sp}^*(n)\oplus \mathfrak{sp}^*(1)$
 \\ \hline
$16$  &  $ \mathfrak{f}(4)$ &  $ \mathfrak{spin}(7)$
  \\ \hline
  $14$  &  $\mathfrak{g}(3) $ & $\mathfrak{g}_2 $
  \\ \hline
  $8$  &  $ \mathfrak{D}(2,1,\alpha)$ &  $ \mathfrak{so}(3)\oplus \mathfrak{so}(3)$
  \\ \hline
  $8$  &  $ \mathfrak{sl}(2\vert 2)/1_{4\times 4}$ &  $ \mathfrak{so}(3)$
  \\ \hline
 \end{tabular}
 \vskip 0.2cm
 \end{center}
\end{table}

\subsubsection{Aspects of geometry}

We summarize the conditions we have obtained so far from imposing the closure of the symmetry superalgebra to simplify
 the relations (\ref{ininin}) and (\ref{conconrs}).  In particular, one finds that
 \bea
i_{Z^{rs}}d\Delta=0~,~~~i_{\check W}d\Delta=0~,~~~ {\tilde \nabla}_{(i} Z_{j)}^{rs}=0~,~~~{\tilde \nabla}_{(i} \check W_{j)}=0~,
\label{inininx}
\eea
and
\bea
&&-2 \langle \phi_+^r, \phi_+^s\rangle-\langle \phi_+^r, \phi_+^s\rangle i_h \check W+2 \langle\Gamma_+\phi^{(r}_-, \Theta_+\phi^{s)}_+\rangle=0~,~~~-i_h Z^{rs}+2 \langle\Gamma_+\phi^{[r}_-, \Theta_+\phi^{s]}_+\rangle=0
\cr
&&  \langle \phi_+^r, \phi_+^s\rangle i_{\check W} (dh)+2 d \langle\Gamma_+\phi^{(r}_-, \Theta_+\phi^{s)}_+\rangle=0~,~~~~i_{Z^{rs}} (dh)+2 d \langle\Gamma_+\phi^{[r}_-, \Theta_+\phi^{s]}_+\rangle=0~,
\cr
&& 2 \langle\Gamma_+\phi^{(r}_-, \Theta_+\phi^{s)}_+\rangle-\Delta \langle \phi_-^{r}, \phi_-^{s}\rangle=0~,~~~
\langle \phi_+^r, \phi_+^s\rangle  \check W + \langle \phi_-^r, \phi_-^s\rangle  h+d \langle \phi_-^r, \phi_-^s\rangle=0~.
\la{conconrsx}
\eea
  It follows that
 \bea
 {\cal L}_{\check W} h=0~,~~~{\cal L}_{Z^{rs}} h=0~.~~~
 \eea
 Moreover, the invariance conditions
of the Killing spinors (\ref{cwphi1}) and (\ref{cwphi2}) imply that
\bea
{\tilde{\cal L}}_{\check W} \phi_\pm^t=0~,~~~2 \langle \phi_+^r, \phi_+^s\rangle \phi_-^t
 -2 \langle \phi_-^r, \phi_-^s\rangle \Gamma_- \Theta_+ \phi_+^t+\langle \phi_+^{r}, \phi_+^{s}\rangle \,{ \slashed{\check W}} \Gamma_-\phi_+^t=0~.
 \label{relphipm3}
 \eea
In particular, the inner products of $\phi_\pm^r$ are invariant under $\check W$.  Notice also that the inner product $\langle \phi_-^{r}, \phi_-^{s}\rangle$ is proportional to $\langle \phi_+^{r}, \phi_+^{s}\rangle$.  For this set $r=s=1$ in the above equation and take the inner product with $\phi_-^{t'}$.  After using the first equation in (\ref{conconrsx}), one can express the inner
product of $\phi_-$'s in terms of that of $\phi_+$'s.

\subsubsection {Superalgebras of AdS$_2$ backgrounds}

The AdS$_2$ backgrounds that preserve $N>2$ supersymmetries are special cases of near horizon geometries for which $\check W=0$. The conditions on the geometry
of internal spaces can be found after setting $\check W=0$ in (\ref{inininx}) and (\ref{conconrsx}).  The symmetry superalgebras are again given in table 1.
One difference is that the vector fields generated by the $\mathfrak{sl}(2, \bR)$ subalgebra are all tangent to the AdS$_2$ subspaces.

Furthermore, setting $\check W=0$ in (\ref{conconrsx}), one finds that $\langle\Gamma_+\phi^{(r}_-, \Theta_+\phi^{s)}_+\rangle=\langle \phi_+^r, \phi_+^s\rangle$ and so
\bea
\langle \phi_-^{r}, \phi_-^{s}\rangle= 2 \Delta^{-1} \langle \phi_+^r, \phi_+^s\rangle~,~~~~h= d\log \Delta^2~,
\eea
where $\Delta$ is no-where vanishing.  The remaining independent conditions on the geometry are that $i_{Z^{rs}} d\Delta=0$ and $Z^{rs}$ are Killing vectors of the internal space.
Also for $\check W=0$, the relation of $\phi_-$ to $\phi_+$ spinors in (\ref{relphipm3}) considerably simplifies.

\subsection{Superalgebras of horizons with $\mathrm{Index}\, \slashed{D}\not=0$}

\subsubsection{$N_-=0$}

So far we have investigated the superalgebras of near horizon geometries and warped AdS$_2$ backgrounds provided  that $\mathrm{Index} \slashed{D}=0$.  Now instead let us take $N_-=0$ and  $\mathrm{Index} \slashed{D}\not=0$ in  (\ref{nindex}).  In this case,
all the Killing spinors are of the form
\bea
\epsilon^s=\phi^s_++ r\Gamma_-\Theta_+\phi^s_+~,~~~s=1,\dots, N~.
\label{ckspixx}
\eea
The only non-vanishing bilinear is $K_{22}^{rs}$ given in (\ref{krsv}).  The spinorial derivatives can be computed in a straightforward way to
reveal that the symmetry superalgebra that arises is  of Poincar\'e type, i.e.~the only non-vanishing (anti-) commutators are
\bea
\{Q^r, Q^s\}= \langle \phi_+^r, \phi_+^s\rangle~ V~,
\label{nilanti}
\eea
where $Q^r$ are the odd generators associated to the Killing spinors (\ref{ckspixx}) and $V$ is an even generator associated to the vector field $\partial_u$.  The symmetry superalgebra
is nilpotent and it can be identified as a supersymmetric version of a  Heisenberg type of algebra.

\subsubsection{$N_-\not=0$ }

It remains to investigate the cases with non-trivial fluxes for which there is at least one Killing spinor $\phi_-$, i.e.~$N_-\not=0$. We shall demonstrate that this case  $\mathrm{Index} \slashed{D}=0$,
and so the symmetry superalgebras of such backgrounds are tabulated in table 1.

To show this, let $\phi_-$ be a Killing spinor. Define  $\phi_+=\Gamma_+\Theta_-\phi_-$  and consider another Killing spinor $\hat\phi_+$
which without loss of generality can be chosen orthogonal to $\phi_+$ and cannot be written as $\hat\phi_+\not=\Gamma_+\Theta_-\hat\phi_-$ for some $\hat\phi_-$.  In this case, there is an $\mathfrak{sl}(2,\bR)$ symmetry generated by the spacetime Killing spinors $\epsilon_A$ in (\ref{kkksp1})
constructed from $\phi_\pm$ as for the  $N=2$ near horizon geometries described in section 2.1.3.

Consider the spacetime Killing spinor $\hat\epsilon_2=\hat\phi_++r \Gamma_-\Theta_+\hat\phi_+$; one finds that
\bea
{\cal L}_{K_{11}} \hat\epsilon_2&=& 2 u\pl\phi_+\pl^2 \hat\epsilon_2+ 2 \pl\phi_-\pl^2 \Gamma_-\Theta_+\hat\phi_+-\slashed{W} \Gamma_-\hat\phi_++ 2u \tilde{\cal L}_W \hat\epsilon_2~,
\eea
where $K_{11}$ is given in (\ref{1fbiln2}).
As the superalgebra must close on  the Killing spinors $\epsilon_1, \epsilon_2$ and $\hat\epsilon_2$ we write ${\cal L}_{K_{11}} \hat\epsilon_2=\alpha \epsilon_1
+\beta \epsilon_2+\gamma \hat\epsilon_2$ for some constants $\alpha, \beta$ and $\gamma$.  This in particular yields
\bea
2 \tilde{{\cal L}}_W\hat\phi_++ 2\pl\phi_+\pl^2 \hat\phi_+=\alpha \phi_+~.
\eea
Taking the inner product with $\hat\phi_+$, one finds that $\pl\phi_+\pl~\pl\hat\phi_+\pl=0$ and so either $\phi_+$ or $\hat\phi_+$ must vanish which
is a contradiction.  This generalizes to backgrounds with non-trivial fluxes and $N_->1$,  and any number of $\hat\epsilon_2$ Killing spinors. We conclude that if there is a Killing spinor $\phi_-$, then for near horizon geometries with non-trivial fluxes all spacetime spinors are generated by pairs $(\phi_+, \phi_-)$ with
$\phi_+=\Gamma_+\Theta_-\phi_-$. Hence the index of the Dirac operator vanishes for all such backgrounds\footnote{The vanishing of the index of the Dirac operator as a consequence of a symmetry argument is reminiscent of the results of \cite{atiyah}.}.  Perhaps this is not unexpected, as the
presence of the $\mathfrak{sl}(2,\bR)$ symmetry pairs up the $\epsilon_1$ and $\epsilon_2$ Killing spinors.

\newsection{Near horizon geometries preserving more than half of supersymmetry}

\subsection{The spatial horizon sections are homogeneous}

The main task of this section is to show that the spatial horizon section ${\cal S}$ of
near horizon geometries preserving more that half of supersymmetry of a theory is a homogeneous space
admitting a transitive action of a group with Lie algebra $\mathfrak{t}_0\oplus \mathfrak{so}(2)$.  The proof relies on an adaptation of the homogeneity theorem of \cite{homogen}.

First observe that as the near horizon geometries are required to preserve more than half of the supersymmetry, at least one of the spacetime Killing spinors is constructed from a $\phi_-$ type spinor
on ${\cal S}$.  Moreover demanding that some of the fluxes do not vanish\footnote{If the fluxes vanish and  the near horizon geometries exit, they  are locally products  $\bR^2\times {\cal S}$, where ${\cal S}$ is
a product of typically special holonomy manifolds that admit parallel spinors. Apart from perhaps  $\bR^2\times T^{D-2}$ all the rest preserve less than half of supersymmetry.  Also
 such solutions  may not be consider as near horizon geometries of black holes.},   we conclude from the results of the previous section that all spacetime Killing spinors are constructed from pairs $(\phi_-, \phi_+)$ with $\phi_+=\Gamma_+\Theta_-\phi_-$
and the symmetry superalgebras are given in table 1.

The Killing vectors on ${\cal S}$ constructed as bilinears can be rewritten as
\bea
W^{rs}=-\langle \phi^r_-, \Gamma^i \Gamma_- \phi^s_+\rangle \tilde \partial_i~,
\eea
and $\phi^s_+=\Gamma_+\Theta_-\phi^s_-$. We shall show that for near horizon geometries that preserve more than half of the supersymmetry, the vectors $W^{rs}$ span the tangent space of ${\cal S}$ at every point $p\in {\cal S}$.

 It suffices to do the
calculation pointwise.  For this define the subspace ${\cal W}_p=\bR\langle W^{rs}\vert_p\rangle\subseteq T_p{\cal S}$ and  ${\cal W}_p^\perp$ its orthogonal
complement in $T_p{\cal S}$. If the $W^{rs}$ do not span $T_p{\cal S}$, then there is a $v\in {\cal W}_p^\perp$, $v\not=0$, such that
\bea
v^r W^{rs}_i\vert_p= -v^i \langle \phi^r_-, \Gamma_i \Gamma_- \phi^s_+\rangle\vert_p= - \langle \phi^r_-, \slashed {v} \Gamma_- \phi^s_+\rangle\vert_p=0~.
\label{kperpk}
\eea
This in particular implies that all the spinors $\slashed {v} \Gamma_- \phi^s_+$ are orthogonal to the Killing spinors $\phi^r_-$. Let   $\mathfrak{S}$ be the spin bundle with sections $\phi_-$. Denoting the span of Killing spinors with ${\cal K}_p=\bR\langle \phi_-^r\vert_p\rangle$ in the fibre $\mathfrak{S}_p$ of $\mathfrak{S}$ over $p$ and ${\cal K}_p^\perp$ its complement,
as the solutions preserve more than half of the supersymmetry, $\mathrm{dim }\,{\cal K}_p >  \mathrm{dim }\,{\cal K}_p^\perp$.

On the other hand  $\slashed {v} \Gamma_- \phi^s_+= \slashed {v}\Gamma_-\Gamma_+ \Theta_-\phi_-^s=2 \slashed {v}  \Theta_-\phi_-^s$.  Moreover it follows from (\ref{kperpk}) that   $\slashed {v}  \Theta_-\vert_p: {\cal K}_p\rightarrow {\cal K}_p^\perp$ .  In all supergravities
for which the horizon conjecture applies and for which the fluxes are non-trivial,    $\mathrm{Ker}\,  \Theta_-=\{0\}$.  Also  $\mathrm{Ker}\, \slashed {v} \Theta_-=\{0\}$ provided\footnote{The signature of the metric of ${\cal S}$ is Euclidean and so the only vector with zero length is the zero vector.} that $v\not=0$ as
   $\slashed {v}^2=v^2 {\bf 1} \not=0$.  So $\slashed {v} \Theta_-$ is an injection, but this is not possible as $\mathrm{dim }\,{\cal K}_p >  \mathrm{dim }\,{\cal K}_p^\perp$.
   This leads to a contradiction unless $v=0$ and so ${\cal W}_p$ spans all $T_p{\cal S}$.  Therefore ${\cal S}$ must be a homogeneous space.
The group $G$ that acts transitively on ${\cal S}$ has Lie algebra $\mathfrak{t}_0\oplus \mathfrak{so}(2)$.  The Killing vector fields generated by $\mathfrak{t}_0\oplus \mathfrak{so}(2)$
are $Z^{rs}$ and $\check W$.  In particular  $\check W$ is generated by the $\mathfrak{so}(2)$ subalgebra.

We remark that for warped $AdS_2\times_w{\cal S}$ backgrounds $\check W=0$.  In such a case, the internal space ${\cal S}$ is homogeneous and the group $G$  that acts transitively on it has Lie algebra $\mathfrak{t}_0$.

\subsection{A no-existence theorem for $N>16$ AdS$_2$ backgrounds and  black hole horizons}

Suppose that the AdS$_2$ backgrounds and the near horizon geometries preserve more than 16 supersymmetries in 10- and 11-dimensional supergravities and have non-trivial fluxes.
In this case the internal space or the spatial horizon section has dimension either 8 or 9 and must be homogeneous.  Furthermore, an inspection
of table 1 reveals that the only possible choices of the  $\mathfrak{t}_0/\mathfrak{c}$ subalgebra are
\bea
&&\mathfrak{so}(N/2)~~(N=18, 20, 22, 24, 26, 28, 30)~;~~~ \mathfrak{u}(N/4)~~(N=20, 24, 28, )~;~~~
\cr
&&\mathfrak{sp}(3)\oplus \mathfrak{sp}(1)~~(N=24)~.
\label{n16alg}
\eea
Near horizon geometries and warped AdS$_2$ backgrounds that preserve 32 supersymmetries have already been excluded in \cite{maxsusy}.
Moreover there are no such backgrounds with 30 supersymmetries in 11-dimensional supergravity \cite{n3011d} and with $\geq 28$ supersymmetries in IIB \cite{n28iib}.

If there are any near horizon geometries  that preserve more than 16 supersymmetries, the near horizon section ${\cal S}$ must admit
a transitive group action with Lie algebra $\mathfrak{t}_0\oplus \mathfrak{so}(2)$, where $\mathfrak{t}_0/\mathfrak{c}$ is given in (\ref{n16alg}).  Similarly,
the internal space of a warped AdS$_2$ background with the most general allowed fluxes that preserves more than 16 supersymmetries  must admit
a transitive group action of a group $G$ with Lie algebra $\mathfrak{t}_0$, where again $\mathfrak{t}_0/\mathfrak{c}$ is given in (\ref{n16alg}).  In both cases $\mathfrak{t}_0/\mathfrak{c}$  must act effectively on either
the spatial horizon section or the internal space.  This is because even generators in the symmetry superalgebra are introduced whenever there is
a non-vanishing vector spinor bilinear leading to a non-trivial action on the spacetime.  Moreover, the even generators of $\mathfrak{t}_0/\mathfrak{c}$ act affectively on $\mathfrak{g}_1$
and so they are not allowed  to vanish.

The relevant 8- and 9-dimensional homogeneous spaces have been classified in \cite{klaus}. An inspection reveals that the only homogeneous spaces that
satisfy these criteria are $S^9=SO(10)/SO(9)$ in 9-dimensions, and $S^8=SO(9)/SO(8)$ and $\bCP^4$ in 8-dimensions.  $\bCP^4$ does not admit
a spin structure so it should be excluded.  The remaining cases can also be excluded.  As $S^9$ is a symmetric space and  does not admit invariant
2-, 3- and 4-forms, if follows that the 4-form flux of 11-dimensional supergravity vanishes. As a result,  the  spacetime field equations cannot be satisfied for non-zero $\Delta$ or $h$.
Potentially, there is an additional case in 11-dimensional supergravity for ${\cal S}=S^8\times S^1$, where $S^1$ may arise due to an element in $\mathfrak{c}$.
However as the relevant superalgebra here is $\mathfrak{osp}(9\vert2)$, $\mathfrak{c}=\{0\}$. Alternatively, one can verify by an explicit computation that there is
no solution with either $\Delta$ or $h$ non-vanishing and  ${\cal S}=S^8\times S^1$.

Similarly,  $S^8$ is  symmetric, and all invariant forms are parallel.
Furthermore, as a consequence of the homogeneity theorem, all scalar fields in IIB and (massive) IIA supergravity must be constant for $N>16$ supersymetries.
An inspection of the IIB Einstein equation in \cite{iibhor} along the spatial
horizon section ${\cal S}$ implies that it must be Ricci flat which is a contradiction as $S^8$ has a non-vanishing Ricci tensor. In (massive) IIA theory, the only non-vanishing flux is associated with the electric component of the 2-form field strength.  An inspection
of the dilatino KSE along ${\cal S}$ in \cite{miiahor} reveals that there are no solutions that preserve $N>16$ supersymmetries unless both this component, and the mass parameter (of the massive theory), vanish. In this case,
the Ricci tensor of ${\cal S}$ vanishes  which is a contradiction as $S^8$ has a non-vanishing Ricci tensor.  Also there are no near horizon geometries with non-trivial fluxes and AdS$_2$ backgrounds in heterotic supergravity that preserve more than 8 supersymmetries \cite{adshet}.

Collecting the results above, we conclude that there are no smooth near horizon geometries with non-trivial fluxes and warped AdS$_2$ backgrounds in 10- and 11-dimensions that preserve more than half of the supersymmetry provided that either the spatial horizon section or the internal manifold, respectively, are compact without boundary.
This also excludes the possibility of constructing solutions by taking discrete identifications.  However if the above assumptions are removed, there are solutions.  For example
the maximally supersymmetric $AdS_4\times S^7$ and $AdS_7\times S^4$ \cite{fr, duffpopemax, townsend} solutions of 11-dimensional supergravity can be re-interpreted as warped AdS$_2$ solutions.  In this
case  the internal space is not compact.

\newsection{Concluding remarks}

We have identified the symmetry superalgebras $\mathfrak{g}$ of near horizon geometries for all theories for which the horizon conjecture applies.  These include all
10- and 11-dimensional supergravity theories as well as ${\cal N}=2$  four- and ${\cal N}=1$ five-dimensional (gauged) supergravities.  We have found that either
the even subalgebra decomposes
as $\mathfrak{g}_0=\mathfrak{sl}(2, \bR)\oplus \mathfrak{t}_0$ and $\mathfrak{t}_0/\mathfrak{c}$ is the Lie algebra of a group acting transitively and effectively
on spheres, where $\mathfrak{c}$ is the center, or $\mathfrak{g}$ is nilpotent.  The latter case arises whenever the number of supersymmetries is equal to the index
 of a Dirac operator.  The symmetry superalgebras can be found in table 1 and in (\ref{nilanti}), respectively.

  The warped AdS$_2$ backgrounds with the most general allowed fluxes are special cases of near horizon geometries for which the internal space is identified with the spatial horizon section and the metric and fields are
  restricted appropriately. Because of this the symmetry superalgebras of warped AdS$_2$ backgrounds are identified in the same way as those of
  near horizon geometries.
The only difference is that for generic near horizon geometries the spatial horizon section admits the action of a group with Lie algebra
$\mathfrak{t}_0\oplus \mathfrak{so}(2)$ while for AdS$_2$ backgrounds the internal space admits the action of a group with Lie algebra $\mathfrak{t}_0$.  The results of this paper together with those in \cite{AdSsuperalgebras} lead to the identification of all symmetry superalgebras of AdS backgrounds in 10- and 11-dimensional supergravities, see also \cite{charles} for results in the AdS$_4$ case.

Furthermore we demonstrated that  there are no near horizon geometries and AdS$_2$ backgrounds that preserve $N>16$ supersymmetries in
10- and 11-dimensional supergravity theories under the same assumptions as those utilized to prove the horizon conjecture. This together with the results of \cite{maxsusy}, \cite{ads5n16} and \cite{ads4n16} classify all the  warped  AdS$_n$ backgrounds, for $n=2,4,5$  in 10- and 11-dimensional
supergravity theories that preserve $N>16$ supersymmetries up to at most discrete identifications.  Product solutions $AdS_n\times M^{D-n}$ with  $M^{D-n}$ symmetric
space, $D=10,11$,   have been classified in \cite{figueroaa}-\cite{Wulffb}.
It is clear from these that there are strong constraints in the existence of AdS backgrounds, especially those that preserve more than half of supersymmetry.   In the context of AdS/CFT, this  implies that
there are very few local geometries that can be identified as gravitational duals of superconformal theories with a large number of supersymmetries.

\setcounter{equation}{0}

\vskip 0.5cm
\noindent{\bf Acknowledgements} \vskip 0.1cm
\noindent   UG is supported by the Swedish Research Council. JG is supported by the STFC Consolidated Grant ST/L000490/1.
GP is partially supported from the  STFC  grant ST/J002798/1.
\vskip 0.5cm

\setcounter{section}{0}
\setcounter{subsection}{0}

\end{document}